\documentclass[aps,prl,superscriptaddress,twocolumn]{revtex4-2}
\usepackage{graphicx} 
\usepackage{subfigure}
\usepackage{multirow}
\usepackage{xspace}
\usepackage{fancyhdr}
\usepackage{amsmath,amssymb}
\usepackage{xcolor}

\usepackage{lineno}

\newcommand{\tc}{$T_{\text{c}}$\xspace}
\newcommand{\pdelta}{{$2\Delta_{\pi}$}\xspace}
\newcommand{\sdelta}{{$2\Delta_{\sigma}$}\xspace}

\newcommand{\kv}{kV/cm\xspace}
\newcommand{\bpf}{$\Omega$\xspace}
\newcommand{\mgb}{MgB$_2$\xspace}

\begin{document}
\title{Amplitude mode in a multi-gap superconductor MgB$_2$ investigated by terahertz two-dimensional coherent spectroscopy}
\author{Kota~Katsumi}
\email{kota.katsumi@nyu.edu}
\affiliation{William H. Miller III Department of Department of Physics and Astronomy, The Johns Hopkins University, Baltimore, Maryland 21218, USA}
\affiliation{Center for Quantum Phenomena, Department of Physics, New York University, New York, New York 10003, USA}
\author{Jiahao~Liang}
\affiliation{William H. Miller III Department of Department of Physics and Astronomy, The Johns Hopkins University, Baltimore, Maryland 21218, USA}
\author{Ralph~Romero III}
\affiliation{William H. Miller III Department of Department of Physics and Astronomy, The Johns Hopkins University, Baltimore, Maryland 21218, USA}
\author{Ke Chen}
\affiliation{Department of Physics, Temple University, Philadelphia, Pennsylvania 19122, USA}
\author{Xiaoxing Xi}
\affiliation{Department of Physics, Temple University, Philadelphia, Pennsylvania 19122, USA}
\author{N.~P.~Armitage}
\affiliation{William H. Miller III Department of Department of Physics and Astronomy, The Johns Hopkins University, Baltimore, Maryland 21218, USA}

\begin{abstract}
We have investigated the terahertz (THz) nonlinear response \textcolor{black}{of the} multi-gap superconductor, \mgb, using THz two-dimensional coherent spectroscopy (THz 2DCS). With broad-band THz drive \textcolor{black}{fields}, we identified a nonlinear response \textcolor{black}{at twice} the lower superconducting gap energy \pdelta at the lowest temperatures. Using narrow-band THz driving pulses, \textcolor{black}{we observed first (FH) and third harmonic responses. The FH intensity shows a monotonic increase with decreasing temperature when properly normalized by the driving field strength}. This is distinct from the single-gap superconductor NbN, where the FH signal exhibited a resonant enhancement at temperatures when \textcolor{black}{twice} the gap energy \textcolor{black}{2$\Delta$} was resonant with the driving photon energy, which \textcolor{black}{was} interpreted to originate from the superconducting amplitude mode. Our results in \mgb are consistent with a well-defined amplitude mode only at the lowest temperatures and indicate strong damping as temperature increases. This likely indicates the importance of interband coupling in \mgb and its influence on the nature of the amplitude mode and its damping.
\end{abstract}

\maketitle
Multidimensional coherent spectroscopy gives new possibilities to acquire information about physical systems, which other spectroscopies cannot~\cite{Cundiff2013}. Recently, it has been used in the terahertz (THz) frequency range, in the form of THz two-dimensional coherent spectroscopy (THz 2DCS)~\cite{Woerner2013}, to explore low-energy electrodynamics in various quantum materials, such as magnons~\cite{Lu2017,mashkovich2021terahertz,blanck2023,zhang2024coupling,zhang2024upconversion}, phonons~\cite{Folpini2017,Blank2023_MBT}, plasmons~\cite{Houver2019,liu2023,Salvador2024,taherian2024squeezed}, ferroelectric soft modes~\cite{pal2021}, correlated metals~\cite{barbalas2025energy} electronic excitations in graphene~\cite{Bowlan2014}, LiNbO$_3$~\cite{Somma2014}, GaAs quantum wells~\cite{Maag2016}, and electronic glasses~\cite{Mahmood2021}. In the case of a conventional superconductor NbN, we have previously demonstrated that the first-harmonic (FH) response of THz 2DCS \textcolor{black}{(the third-order nonlinear response appearing at the same frequency as the drive $\Omega/2\pi$)} is very sensitive to the amplitude mode of the superconducting (SC) order parameter via its paramagnetic light-matter coupling~\cite{Katsumi_NbN_2023}. Among other aspects, it showed a resonant enhancement at temperatures near \tc when the \textcolor{black}{amplitude-mode} energy coincided with the THz driving photon energy.

The THz 2DCS response in a multi-gap superconductor is of particular interest, as it may host at least two amplitude modes and a relative phase mode of its order parameters, namely the Leggett mode~\cite{leggett1966,Blumberg2007,Anishchanka2007,Klein2010,Cea2016}. \mgb is known to be a $s$-wave multi-gap system~\cite{Xi2008} and is an ideal material to investigate this physics.  THz nonlinear responses have been reported in \mgb using pump-probe spectroscopy~\cite{Giorgianni2019,Yuan2024selective} or THz third-harmonic generations (THG)~\cite{Kovalev2021,Reinhoffer2022}. Theoretical work has shown that in dirty-limit superconductors, the paramagnetic light-matter coupling plays a more crucial role in the THz nonlinear responses than the diamagnetic one, and the respective contributions of the amplitude modes, BCS quasiparticle excitations, and the Leggett mode depend on the level of disorder~\cite{Murotani2019, Jacopo2022}. It has also been pointed out that the amplitude modes in multi-gap superconductors can exhibit marked differences compared to those in single-band counterparts due to \textcolor{black}{their} sensitivity to inter-band couplings~\cite{Jacopo2022}. Although the THz range non-linearities have been investigated previously in multi-gap systems~\cite{Luo2023}, the low-drive field limit where intrinsic \textcolor{black}{SC} properties are relevant was not emphasized.

To investigate the nature of the amplitude mode in a multi-gap superconductor, we performed THz 2DCS on a \mgb thin film, with \tc~=~38~K. The details of the sample properties and the experimental setup are described in the Supplementary Material (SM)~\cite{sm}. Using broad-band THz pulses, we identified a nonlinear response at \textcolor{black}{twice} the lower SC gap energy \pdelta at the lowest temperatures in the SC state. To further resolve the spectral features of the nonlinear response, we employed narrow-band THz pulses at the frequency of $\Omega/2\pi$~=~0.63~THz as the drive. Below \tc, we identified a nonlinear signal at FH \textcolor{black}{($\Omega$)} and third-harmonic \textcolor{black}{(TH, $3\Omega$)} frequencies. 
When the FH intensity is normalized by the THz field strength inside the film, it shows a monotonic increase with decreasing temperature.  This is distinct from single-gap NbN where the normalized FH intensity is resonantly enhanced when twice the SC gap matches $\Omega$~\cite{Katsumi_NbN_2023}. We can fit the FH temperature dependence with a model that assumes an over-damped amplitude mode at $\Omega=2\Delta_\pi$. Our results indicate that the essential difference in the nonlinear response between \mgb and NbN, is likely due to the effect of interband coupling, which modifies the line width of the amplitude mode in the $\pi$ band.

\begin{figure}[t]
	\centering
	\includegraphics[width=\columnwidth]{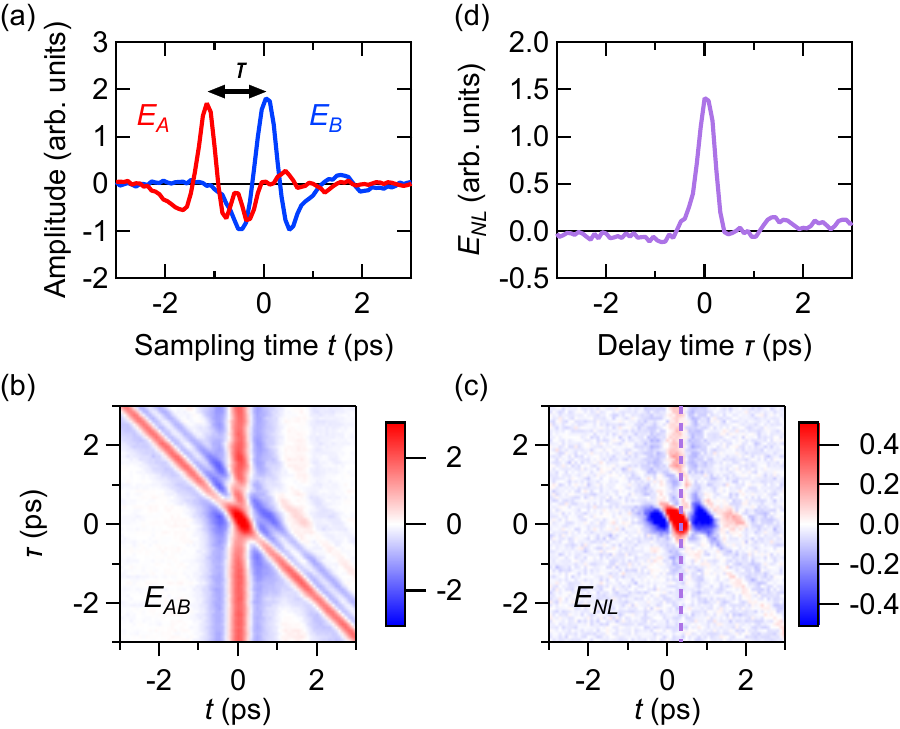}
	\caption{(a) Time traces of the $A$ and $B$ pules transmitted through the \mgb sample as a function of the sampling time~$t$. The delay time between the $A$ and $B$ pulses is denoted as~$\tau$. (b) 2D plot of the time traces of the $A$ and $B$ pulses together ($E_{AB}(t,\tau)$) transmitted after the sample at 7~K as a function of $t$ and $\tau$. (c) The nonlinear difference signal $E_{NL}(t,\tau)$ from data in (b). (d) Time evolution of the nonlinear signal $E_{NL}(t,\tau)$ as a function of $\tau$ at a fixed sampling time $t=$\textcolor{black}{0.36}~ps, indicated by the vertical dashed line in (c).}
	\label{fig1}
\end{figure}

To perform THz 2DCS, we generated two intense broad-band THz pulses by the tilted-pulse front technique with two LiNbO$_3$ crystals~\cite{Hebling2002,Watanabe2011,Hirori2011} (see SM for details). The time traces of the two THz pulses are measured by sweeping the timing \textcolor{black}{$t$} between the sampling pulse and the THz pulses, as shown in Fig.~1(a). We sweep the arrival time of the $A$-pulse with respect to that of the $B$-pulse to shift the delay time $\tau$ between $A$ and $B$ pulses. We measured three sets of the transmitted THz electric fields ($E$-fields): only the $A$-pulse $E_A(t,\tau)$, only the $B$-pulse $E_B(t)$, and both pulses together $E_{AB}(t,\tau)$, \textcolor{black}{which is shown in Fig. 1(b)}.  We obtain the nonlinear signal $E$-field as $E_{NL}(t,\tau) =E_{AB}(t,\tau) - E_{A}(t,\tau) - E_{B}(t,\tau)$, as shown in Fig.~1(c). To obtain the intrisic response it is  essential to set the peak $E$-fields of $A$ and $B$ pulses to less than \textcolor{black}{21}~\kv to prevent the suppression of the SC state.  For fields larger than \textcolor{black}{21}~\kv (Fig.~S3.) data at 7~K  displays a long-lived component, which is reasonably ascribed to the quasiparticle excitations of broken Cooper pairs~\cite{Demsar2003,Giorgianni2019}. Such behavior can be contrasted with the \textcolor{black}{21}~\kv data in Fig.~\ref{fig1}(d) where this long lived component is absent.  \textcolor{black}{It is also worth mentioning that an important aspect of THz 2DCS with its strong probe compared to typical ``pump-probe" schemes  is that the rephasing and non-rephasing signals are relatively enhanced~\cite{Katsumi_NbN_2023}.}




\begin{figure}[t]
	\centering
	\includegraphics[width=\columnwidth]{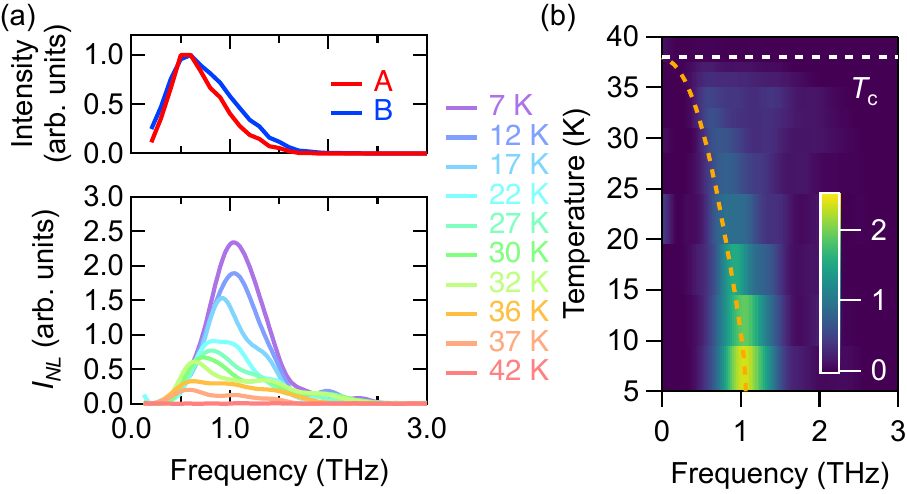}
	\caption{(a)  Top panel presents the power spectrum of the $A$ and $B$ pulses measured at the sample position with a GaP \textcolor{black}{crystal}.  Bottom panel shows the power spectrum of the nonlinear signal at 7~K measured at $\tau=0$~ps with the broad-band THz pulses. (b) The power spectrum of the nonlinear signal as a function of temperature. The orange dashed curve is \pdelta computed by numerically solving the two-band BCS gap equations.}
	\label{fig2}
\end{figure}

In the upper panel of Fig. 2(a), we show the \textcolor{black}{power spectrum of the THz} drive fields at the sample position with no sample in place.  In the lower panel of Fig. 2(a), we show the transmitted nonlinear intensity through the \mgb film. \textcolor{black}{We identify a} peak at 1~THz at 7K, which coincides with the experimentally measured low temperature \pdelta value from the \textcolor{black}{THz optical conductivity}. With increasing temperature, the nonlinear signal's spectral weight displays a redshift, loses intensity, and broadens \textcolor{black}{greatly} at temperatures still well below \tc. Figure 2(b) shows the temperature dependence of the nonlinear signal's power spectra. The peak in the nonlinear signal initially follows the expected dependence for a mean-field two-gap model for \pdelta as shown by the orange dashed curve (see SM for further details), but is lost well before \tc. This comes from an intrinsic spectral property of the material because the peak position does not match the spectral maxima of the driving $A$ and $B$ pulses (shown in the top panel of Fig.~2(a)) nor the shape of the transmission \textcolor{black}{power} spectra presented in Fig.~S1(c) in SM. \textcolor{black}{The nonlinear signal's peak at \pdelta around the lowest temperatures is quite similar to the inferred signatures of the amplitude mode in NbN.} \textcolor{black}{We further consider the effect of the THz transmission coefficient on this resonance lineshape in SM.} However, in broad-band THz 2DCS experiments, multiple difference frequency components can contribute to four-wave mixing processes, complicating interpretation. Therefore, evaluating the precise magnitude and width of the peak can be difficult because normalization of the nonlinear signal by the broad-band $A$ and $B$ pulses is challenging.

\begin{figure}[t]
	\centering
	\includegraphics[width=\columnwidth]{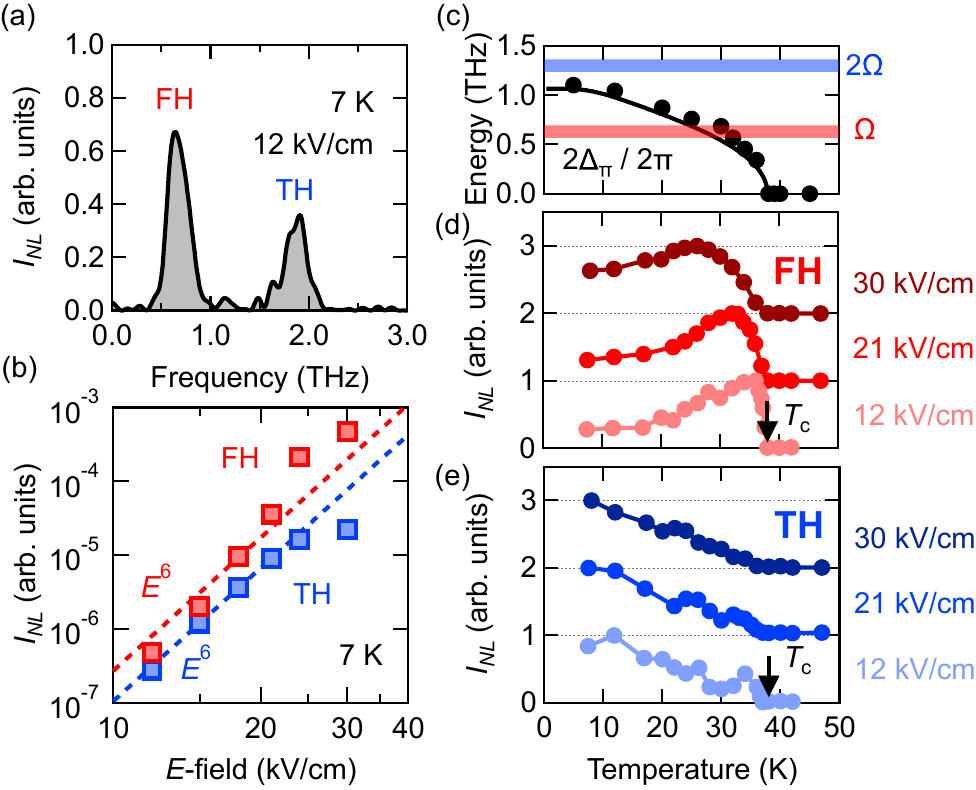}
	\caption{(a) Power spectrum of the nonlinear signal at 7~K measured at $\tau=0$~ps with the narrow-band THz pulses. (b) The frequency-integrated intensity of the FH and TH contributions at 7~K when $\tau=0$~ps as a function of the $B$-pulse peak $E$-field in a log-log plot. Here, both $A$   and $B$ pulses are controlled equally. The dashed curves are the guides to the eye with a slope of 6. (c) Temperature dependence of \textcolor{black}{twice} the SC gap energies \pdelta evaluated from the equilibrium THz optical conductivity (circles). The solid curve is \pdelta computed by numerically solving the two-band BCS gap equations. (d) Raw data of the frequency-integrated intensity of the FH contribution as a function of temperature. The numbers on the right denote the peak $E$-field strength of the driving THz pulses. (e) The same plot as (c) but for the TH contribution.}
	\label{fig3}
\end{figure}

  To further resolve the spectral features of the nonlinear signal, we performed THz 2DCS with a pair of narrow-band THz pulses, which significantly restricts the allowed four-wave mixing process and simplifies the analysis. \textcolor{black}{We specifically chose $\Omega/2\pi$~=~0.63~THz as the driving frequency to focus on investigating the amplitude mode in the $\pi$ band in \mgb, and compare the results with those in the single gap counterpart NbN~\cite{Katsumi_NbN_2023}}.  In Fig.~3(a), we show \textcolor{black}{the measured raw (i.e., unnormalized)} power spectrum of the nonlinear signal from \mgb at 7~K using THz peak fields of 12~\kv with $\tau=0$~ps. The raw spectrum exhibits two peaks at $\Omega/2\pi$~=~0.63~THz and $3\Omega/2\pi$~=~1.9~THz, corresponding to the FH and TH contributions, respectively. The FH and TH intensities follow $E^6$ as shown in Fig.~3(b) up to \textcolor{black}{21} and 25~\kv, respectively, indicating that both are third-order nonlinear responses.

In Figs.~3(d) and (e), we plot the frequency-integrated intensity $I_{NL}$ of the raw FH and TH signals as a function of temperature with the multi-cycle THz pulses when $\tau=0$~ps. Here, the nonlinear signal is integrated from 0.3 to 1~THz for the FH signal and from 1.6 to 2.2~THz for the TH signal.  As shown in Fig.~3(d), for all measured THz $E$-fields, the \textcolor{black}{unnormalized} FH signal displays a peak at a temperature below \tc, but note that unlike NbN it is not found at temperatures where \textcolor{black}{twice} the lower gap satisfies $\Omega = 2\Delta_{\pi}$ (Fig.~3(c)).  Moreover, \textcolor{black}{at larger THz $E$-field strength (particularly for the 25~\kv data)}, the peak temperature decreases as superconductivity is suppressed. This \textcolor{black}{again} highlights the importance of using small drive fields to measure the intrinsic \textcolor{black}{SC} properties. The temperature evolution of the TH signal agrees well with the previous THG experiments, where it \textcolor{black}{reaches} its maximum toward $2\Omega = 2\Delta_{\pi}$~\cite{Kovalev2021}.

In order to consider the actual temperature dependence of any resonant features, one must take into account the screening effect on the THz $E$-field \textcolor{black}{inside the sample}. As presented in Fig.~S4(a) in SM, the transmitted $A$ and $B$ fields from the sample decrease when the temperature is lowered below \tc \textcolor{black}{due to enhanced conductivity.}  Following the procedure in Ref.~\cite{Katsumi_NbN_2023}, we normalize the FH and TH intensities by \textcolor{black}{the sixth power of} the transmitted $E$-field of the $A$-pulse, as shown in Fig.~4. While the \textcolor{black}{qualitative} behavior of the TH signal is unchanged by the normalization, surprisingly, the normalized FH signal shows a monotonic increase when the temperature is decreased for all the driving THz $E$-fields studied here. This is in stark contrast to NbN, whose normalized FH signal exhibits a resonant enhancement when the driving frequency $\Omega$ matches \textcolor{black}{twice} the SC gap $2\Delta$~\cite{Katsumi_NbN_2023}. The same result is obtained when the $B$-pulse is used for normalization (see SM). We stress that one must properly normalize the intensity of the nonlinear response to obtain its correct temperature dependence, as pointed out in previous THG \cite{Katsumi2023,Chu2020} and THz 2DCS experiments \cite{Katsumi_NbN_2023}.

\begin{figure}[t]
	\centering
	\includegraphics[width=\columnwidth]{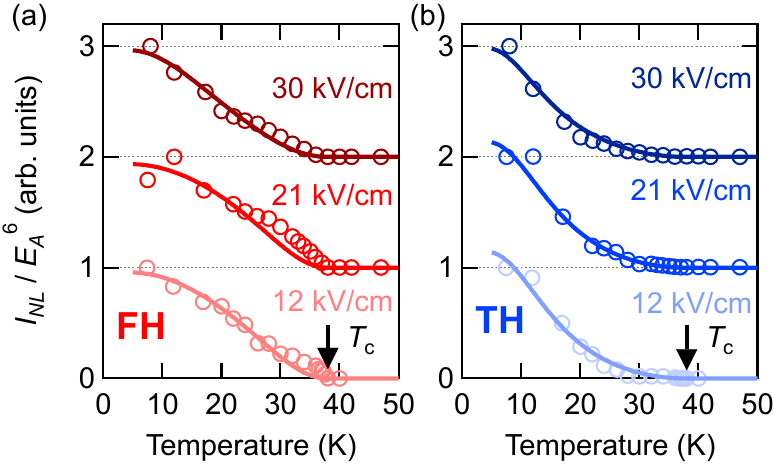}
	\caption{Temperature dependence of the frequency-integrated intensity of the (a) FH and (b) TH contributions normalized by the transmitted $E$-field of the $A$-pulse. \textcolor{black}{Solid} curves are the fits using Eq. \eqref{eq1} with $\omega = \Omega$ for the FH and $\omega = 2\Omega$ for the TH signals.}
	\label{fig4}
\end{figure} 

This indicates that the details of the FH signal in \mgb is distinct from that in NbN, even though their linear responses are similar~\cite{Katsumi_NbN_2023}. In Ref.~\cite{Katsumi_NbN_2023}, it was shown that the FH nonlinear response in NbN is dominated by the amplitude mode for a wide range of impurity levels $3 \leq k_Fl \leq 30$, where $k_F$ is the Fermi wave vector and $l$ is the electronic mean free path. In the case of this \mgb, the zero-frequency resistivity is approximately 50~$\mu\Omega\ $cm and \textcolor{black}{we can estimate $k_Fl$ to be} approximately 13.5~\cite{Sologubenko2002,lal2009}. Therefore, \textcolor{black}{it is reasonable to expect that} the FH nonlinear signal \textcolor{black}{in \mgb} is dominated by amplitude modes as well. \textcolor{black}{Furthermore, since \mgb is cleaner than NbN, the damping of the FH nonlinear signal in \mgb is very unlikely due to the effect of impurity scattering.} The difference in the amplitude modes in single and multi-gap superconductors was theoretically examined in Ref.~\cite{Jacopo2022}, and it was found that the respective ratio of the modes that derive from different bands significantly changes if the interband coupling strength is finite. For the parameters assumed in \mgb, the amplitude mode of the $\pi$ band was strongly suppressed, and that of the $\sigma$ band is pronounced. Nevertheless, we did not observe the resonance when \textcolor{black}{twice} the higher SC gap energy \sdelta matched with $\Omega$ \textcolor{black}{either}. This is \textcolor{black}{likely} because the calculation assumes zero-temperature, while \textcolor{black}{the temperature that satisfies the resonance conditions for the $\sigma$ band $\Omega = 2\Delta_{\sigma}$ is} just below \tc, and the damping is more significant at such a high temperature.

To quantitatively examine the interpretation above, we evaluated the temperature dependence of the FH and TH nonlinear signals by fitting them with a model for the amplitude-mode resonance \textcolor{black}{that was} developed in Ref.~\cite{Katsumi_NbN_2023}:
\begin{align}
	I(\omega,T)= I_0 \frac{\Delta_{\pi}(T)^2}{(\omega+i\delta)^2-(2\Delta_{\pi}(T))^2}.
	\label{eq1}
\end{align}
Here, $\omega$ is the angular frequency, $T$ is the temperature, $\delta$ is the damping rate of the resonance, and $I_0$ is a constant. \textcolor{black}{While incorporating the interband coupling in Eq.~(1) is challenging, it is reasonable to assume that the $\pi$-band amplitude mode has a larger contribution because it is closer to the driving frequency compared to the $\sigma$-band mode.} Given the fact that the broad-band nonlinear signal displays the peak at \pdelta at the lowest temperature, we set the resonant condition in the same manner as NbN, i.e., $\omega=\Omega$ for the FH and $\omega=2\Omega$ for the TH. For the driving field strength of 12~\kv, the FH and TH signal can be simultaneously fit (Fig. \ref{fig4}) with $\delta = 0.55$~THz. \textcolor{black}{The data at \textcolor{black}{21} and 30 \kv are fitted using the same value of $\delta$, whereas \pdelta is reduced by a factor of 0.81 at 30 \kv.} \textcolor{black}{The} obtained $\delta$ is critically larger than the $\delta = 0.12$~THz found in NbN~\cite{Katsumi_NbN_2023}. This result is consistent with the interpretation that the resonance at $\Omega=2\Delta_{\pi}$ in \mgb is strongly damped due to the interband coupling near the inferred resonance temperature. Larger damping could also occur because the temperature where the resonance condition is met is approximately two times higher in \mgb.

In summary, we performed THz 2DCS on the multi-gap superconductor \mgb. Utilizing broad-band THz pulses, we observed a nonlinear signal peaked at \textcolor{black}{twice} the lower SC gap energy $2\Delta_{\pi}$ at our lowest measured temperatures. The signal rapidly decays with increasing temperature. With narrow-band THz pulses, a nonlinear signal is found at the FH and TH of the driving THz frequency \bpf. The normalized FH signal displays a monotonic increase with lowering the temperature, unlike the resonant enhancement found near \tc at $\Omega=2\Delta$ reported in NbN.   This is consistent with a much larger damping rate of the amplitude mode in \mgb, at least in the temperature range of interest where the resonant condition is met. Our observations demonstrate the difference in the SC amplitude mode in \mgb and NbN, and likely highlight the importance of the interband couplings in the SC collective excitations.

We thank L. Benfatto, J. Fiore, M. Udina, and G. Seibold for fruitful discussions. At JHU this project was supported by the Gordon and Betty Moore Foundation, EPiQS initiative, Grant No. GBMF-9454 and NSF-DMR 2226666. K.K. had additional support from an Overseas Research Fellowship of the JSPS. \textcolor{black}{K.C. and X.X. were supported by the U.S. Department of Energy, Office of Science under Grant DE-SC0022330.}

\bibliography{MgB2.bib}
\bibliographystyle{apsrev4-2}
\end{document}


\title{Amplitude mode in a multi-gap superconductor MgB$_2$ investigated by terahertz two-dimensional coherent spectroscopy\\ Supplemental Material}
\author{Kota~Katsumi}
\email{kota.katsumi@nyu.edu}
\affiliation{William H. Miller III Department of Department of Physics and Astronomy, The Johns Hopkins University, Baltimore, Maryland 21218, USA}
\affiliation{Center for Quantum Phenomena, Department of Physics, New York University, New York, New York 10003, USA}
\author{Jiahao~Liang}
\affiliation{William H. Miller III Department of Department of Physics and Astronomy, The Johns Hopkins University, Baltimore, Maryland 21218, USA}
\author{Ralph~Romero III}
\affiliation{William H. Miller III Department of Department of Physics and Astronomy, The Johns Hopkins University, Baltimore, Maryland 21218, USA}
\author{Ke Chen}
\affiliation{Department of Physics, Temple University, Philadelphia, Pennsylvania 19122, USA}
\author{Xiaoxing Xi}
\affiliation{Department of Physics, Temple University, Philadelphia, Pennsylvania 19122, USA}
\author{N.~P.~Armitage}
\affiliation{William H. Miller III Department of Department of Physics and Astronomy, The Johns Hopkins University, Baltimore, Maryland 21218, USA}
\maketitle
\section{Sample details}
High-quality 20 nm-thick \mgb films were grown on a single-crystal MgO (111) substrate by hybrid physical-chemical vapor deposition (HPCVD)~\cite{Xi2007}. The details of the sample fabrication are documented in Refs.~\cite{Xi2007,Giorgianni2019}.

\begin{figure}[t]
	\centering
	\includegraphics[width=\columnwidth]{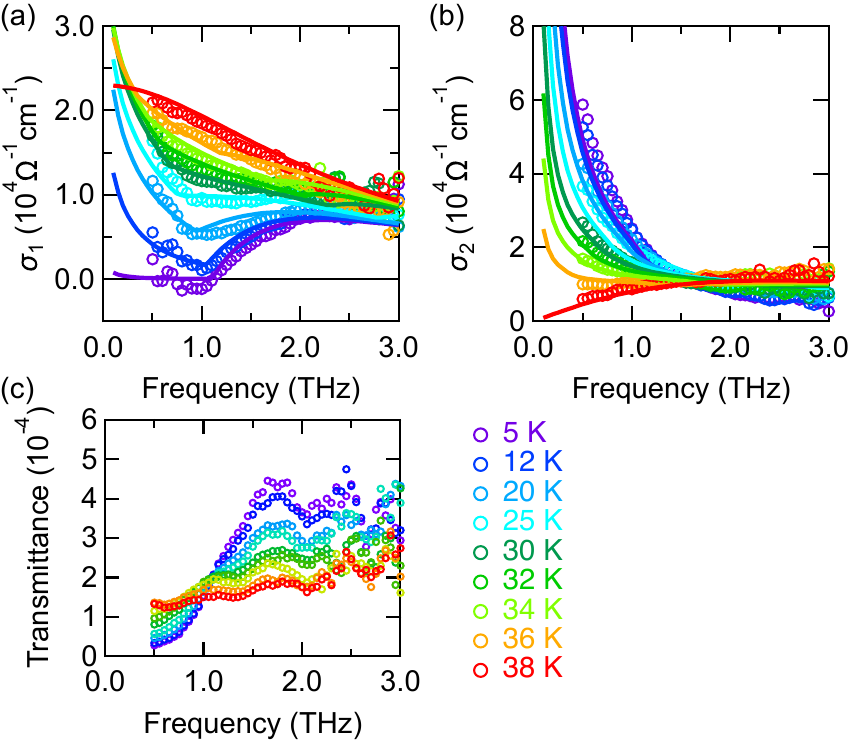}
	\caption{(a) Real and (b) imaginary parts of the optical conductivity of MgB$_2$ (\tc~=~38 K). The solid curves are the optical conductivity calculated by the two-components MB model. (c) THz power transmittance as a function of temperature.}
	\label{fig_s1}
\end{figure}

We evaluated the lower superconducting (SC) gap of the \mgb sample by terahertz (THz) time-domain spectroscopy. The obtained optical conductivity $\sigma(\omega)$ is presented in Fig.~\ref{fig_s1}. In its real part at 5~K, the opening of twice the lower gap \pdelta is identified around 1~THz. We evaluated twice the SC gap \pdelta at different temperatures by the two-component Mattis-Bardeen (MB) model~\cite{Mattis1958,Zimmermann1991,Jacopo2022}. Here, we set the parameters as summarized in Table 1, similar to the values in Ref.~\cite{Jacopo2022}. The values of twice the higher gap \sdelta are adopted from the solutions of the BCS gap equations for two bands as described in the next section. 

We plot the obtained $2\Delta_{\pi}$ as a function of temperature in Fig. \ref{fig_s2} by the red circles. It follows the solutions for \pdelta of the BCS gap equations and is consistent with the previous THz experiments~\cite{Kaindl2001,Kovalev2021} and scanning tunneling microscopy (STM)~\cite{Iavarone2002}.

\begin{table}[h]
	\centering
		\caption{Parameters for the MB model for each band. $2\Delta (T = 0 \ \text{K})$, $\omega_p$, and $\tau_p$ denote twice the SC gap energy at $T$~=~0~K, plasma frequency, and the scattering rate, respectively. $\omega_p$ and $\tau_p$ are kept fixed through all the temperatures.   From Ref.~\cite{Jacopo2022}.}
	\begin{tabular}{ |c| c| c|c| }
		\hline
		\ & \ $2\Delta (T = 0 \ \text{K})/2\pi $ (THz) \ &\  $\omega_p$ (eV)\ & \ $1/\tau_p$ \  (meV)\\
		\hline
		$\pi$ band \ & 1.1 & 5.2 & 12.5  \\
		\hline
		$\sigma$ band \ & 3.4 & 4.2 & 8.33 \\
		\hline
	\end{tabular}
	\label{tab1}
\end{table}

\section{Evaluating the SC gap energies}
We evaluated the temperature dependence of the SC gap energies \pdelta and \sdelta by self-consistently solving the BCS gap equations for two bands~\cite{Kuzmichev2014,Giorgianni2019}:
\begin{align}
	\Delta_\alpha=\sum_{\beta=\pi,\sigma}V_{\alpha\beta}\Delta_{\beta}F(\Delta_\beta),
	\label{eqs1}
\end{align}
where $F(\Delta_\beta)$ is defined as
\begin{align}
	F(\Delta_\beta) = N_\beta\int_{0}^{\omega_c}\text{d}\xi\frac{\tanh(\sqrt{\xi^2+{\Delta_\beta}^2} / 2k_BT)}{\sqrt{\xi^2+{\Delta_\beta}^2}}.
	\label{eqs2}
\end{align}
Here, $\Delta_\beta$ and $N_\beta$ are the SC gap energy and the density of states at the Fermi energy of the $\beta$-th band ($\beta = \pi,\sigma$), respectively. $V_{\alpha\beta}$ are the matrix elements of the interaction, which determine the intraband ($\alpha=\beta$) and interband ($\alpha \neq \beta$) coupling constants, $\omega_c$ is the cut-off frequency, $k_B$ is the Boltzmann constant, $T$ is the temperature. For the case of \mgb, the following dimensionless parameters are believed to describe the SC gap well~\cite{liu2001}: $\lambda_{11}=V_{11}N_1=0.28$, $\lambda_{22}=V_{22}N_2=0.96$, and $\lambda_{12}=\lambda_{21}=V_{12}\sqrt{N_1N_2}=0.19$, and $N_2/N_1=0.73$. We adopted the cut-off frequency as $\omega_c = 1.78$~THz from the literature~\cite{xi2008}.

\begin{figure}[t]
	\centering
	\includegraphics[width=\columnwidth]{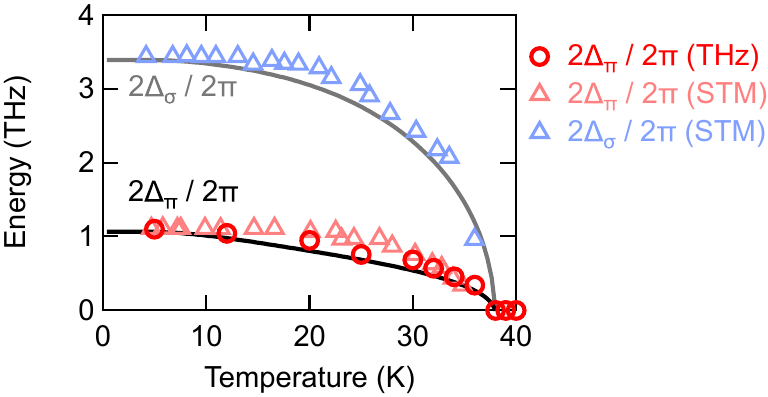}
	\caption{(a) Temperature dependence of the SC gap energy of $\pi$ and $\sigma$ bands of \mgb. Red circles show \pdelta evaluated by  THz time-domain spectroscopy in equilibrium. Red and blue triangles are values of \pdelta and \sdelta adopted from the previous STM experiments~\cite{Iavarone2002}. The solid black and gray curves are the solutions of the BCS gap equations.}
	\label{fig_s2}
\end{figure}

We can solve Eqs.~\eqref{eqs1} for the $\pi$ and $\sigma$ bands self-consistently, and the obtained SC gap energies are presented in Fig. \ref{fig_s2} by the solid curves. The curves show good agreement with the experimental data of \pdelta in our THz experiments (the red circles) and the STM (\pdelta and \sdelta by the red and blue triangles, respectively)~\cite{Iavarone2002}.

\section{Experimental details}
To perform THz two-dimensional coherent spectroscopy (2DCS), we utilized a regenerative amplified Ti:sapphire laser with a center wavelength of 800~nm, pulse duration of 100~fs, pulse energy of 9~mJ, and repetition rate of 1~kHz. We divided the output from the regenerative amplifier into three beams: two for generating intense THz pulses  $A$ and $B$ pulses), and the other for the gate pulse of the electro-optic (EO) sampling. We combined the two THz pulses to the same optical path using a silicon beam splitter. The transmitted THz pulses from the sample were detected by the EO sampling in a 1-mm-thick ZnTe (110) crystal. The obtained peak THz electric-field ($E$-field) strengths for the broad-band $A$ and $B$ pulses could be 150~\kv and 80~\kv, respectively.

For the THz 2DCS experiments, the $A$-pulse's intensity is attenuated to be equal to that of the $B$-pulse. Narrow-band THz pulses at the frequency of $\Omega=0.63$~THz were generated by placing two bandpass filters in the beam pass of the $A$-pulse and one filter for the $B$-pulse. We obtained the peak THz $E$-fields of  30~\kv for both pulses.

\section{THz pump-THz probe experiments}
\begin{figure}[t]
	\centering
	\includegraphics[width=\columnwidth]{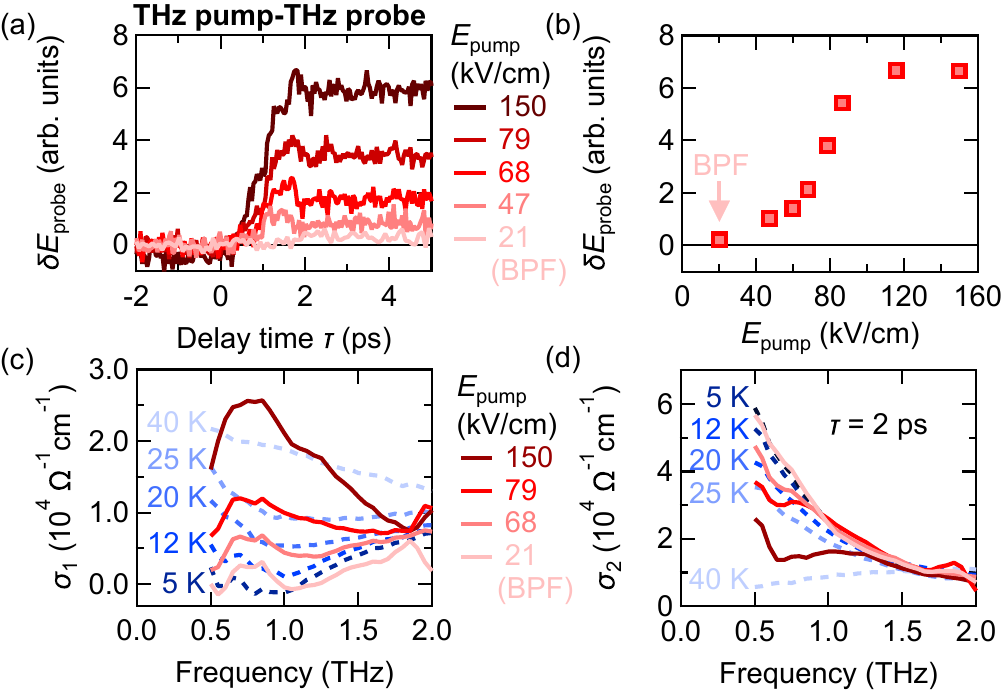}
	\caption{(a) THz pump-induced change in the THz probe $E$-field amplitude ($\delta E_{\text{probe}}$) as a function of the pump-probe delay time \textcolor{black}{$\tau$} measured at 7~K. The peak THz pump $E$-field amplitude is shown in the legend. (b) The maximum value of $\delta E_{\text{probe}}$ as a function of the pump peak $E$-field at 7~K. (c) Real and (d) imaginary parts of the optical conductivity of MgB$_2$ at \textcolor{black}{$\tau=2$~ps} measured at 7~K(solid curves). The dashed curves denote the optical conductivity in equilibrium at selected temperatures.}
	\label{fig_s3}
\end{figure}

We also performed THz pump-THz probe experiments on \mgb at temperatures as low as 7~K to confirm that the SC condensate is not depleted in the THz 2DCS. We employed exactly the same setup as for the THz 2DCS, except for using a weak broad-band probe THz pulse whose peak $E$-field was set to 10~\kv. Figure~\ref{fig_s3}(a) shows the THz pump-induced change in the probe amplitude ($\delta E_{\text{probe}}$) as a function of the pump-probe delay time \textcolor{black}{$\tau$} at the fixed sampling time \textcolor{black}{$t=0.36$~ps}. With the strongest excitation of 150~\kv, we identify a long-lived relaxation of SC quasiparticles. Importantly, $\delta E_{\text{probe}}$ is almost undetectable for the narrow-band pump with BPF. This is clearly seen by plotting the maximum $\delta E_{\text{probe}}$ as a function of the pump peak $E$-field in Fig.~\ref{fig_s3}(b).

We further evaluated the transient optical conductivity after the THz excitation. Here, we obtained the probe time trace by sweeping the arrival time of the probe pulse while keeping the pump and sampling pulses' time fixed~\cite{Isoyama2021,Katsumi2023}. Figure~\ref{fig_s3}(c) compares the real part of the optical conductivity spectra $\sigma_1(\omega)$ out of equilibrium at 7~K (the solid curves) and in equilibrium at selected temperatures (the dashed curve). The corresponding imaginary parts $\sigma_2(\omega)$ are shown in Fig.~\ref{fig_s3}(d). When the pump peak $E$-field is set to 150~\kv, $\sigma_1(\omega)$ measured at \textcolor{black}{$\tau$}~=~2~ps is close to that at 40~K in equilibrium normal state. $\sigma_2(\omega)$ is also strongly suppressed but larger that that at 40~K. These results indicate that the SC condensate is significantly depleted with the THz peak $E$-field of 150~\kv. This result clearly demonstrates that the THz pump-probe or 2DCS with THz field stronger or equal to 150~\kv is depleting the SC state, and the observed nonlinear signals come from the SC quasiparticle excitation. On the other hand, when the sample is driven by the narrow-band THz pump, both $\sigma_1(\omega)$ and $\sigma_2(\omega)$ remain almost intact at \textcolor{black}{$\tau$}~=~2~ps. This results ensures that the SC condensate is not depleted in the  current study of THz 2DCS.

\begin{figure}[t]
	\centering
	\includegraphics[width=\columnwidth]{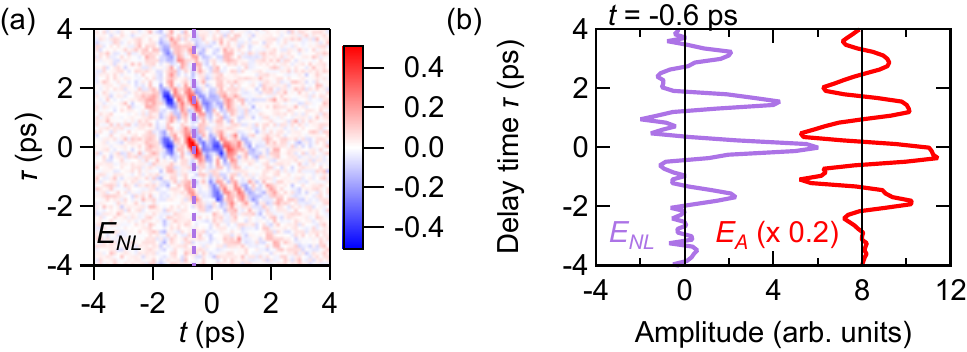}
	\caption{\textcolor{black}{(a) 2D plot of the time traces of $E_{NL}(t, \tau)$ at 7~K as a function of $t$ and $\tau$ using narrow band THz pulses. The driving fields are set to 12~\kv. (b) $E_{NL}(t, \tau)$ and $E_{A}(t, \tau)$ as a function of $\tau$ at a fixed sampling time $t$ = -0.6 ps, indicated by the vertical dashed line in (a).}}
	\label{fig_narrow}
\end{figure}

\textcolor{black}{\section{THz 2D power spectra}}
\textcolor{black}{\subsection{Narrow-band THz pulses}}

\begin{figure}[t]
	\centering
	\includegraphics[width=5cm]{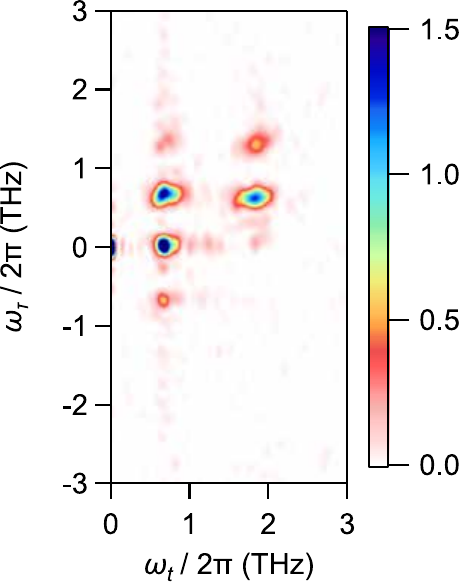}
	\caption{\textcolor{black}{2D power spectra of the nonlinear signal measured at 7~K with the narrow-band THz pulses for $\tau \geq 0$~ps (the $AB$ pulse sequence).}}
	\label{fig_narrow_2D}
\end{figure}

\textcolor{black}{We present the nonlinear signal $E_{NL}(t, \tau)$ at 7~K with the narrow-band driving fields and its cut at $t$ = -0.6 ps in Fig. \ref{fig_narrow}. The nonlinear signal is dominated by the oscillatory behavior.  The long-lived signal that we believe orginates in quasiparticle excitation is hardly seen at this drive field of 12 kV/cm. The power spectrum of $E_{NL}(t, \tau)$ for $\tau \geq 0$ ps (the $A$ pulse arrives earlier than the $B$ pulse) is shown in Fig. \ref{fig_narrow_2D} and exhibits peaks at $(\omega_t, \omega_{\tau}) = (\Omega, \Omega)$ and $(\Omega, -\Omega)$, referred to the non-rephasing and rephasing signals, respectively \cite{Mahmood2021,Katsumi_NbN_2023}. These signals are essentially inaccessible by the usual pump-probe spectroscopy because their amplitudes are proportional to the square of the amplitude of the $B$ (probe) pulse, which in pump-probe schemes is much weaker than that of the $A$ (pump) pulse.}

\begin{figure}[t]
	\centering
	\includegraphics[width=4.5cm]{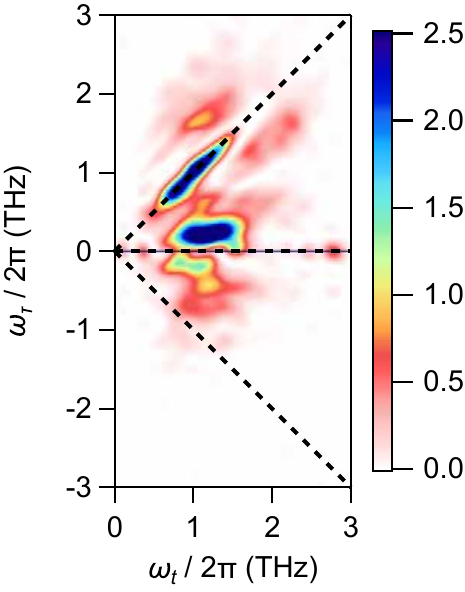}
	\caption{\textcolor{black}{2D power spectra of the nonlinear signal measured at 7~K with the broad-band THz pulses.}}
	\label{fig_broad}
\end{figure}

\textcolor{black}{\subsection{Broad-band THz pulses}}
\textcolor{black}{
It was theoretically proposed that THz 2DCS can disentangle the homogeneous and inhomogeneous broadening, such as in the 1D Ising chain system~\cite{Wan2019}. This concept has been experimentally demonstrated in the cases of disordered silicon~\cite{Mahmood2021} and Josephson plasmon in a $c$-axis cuprate superconductor~\cite{Liu2023,Salvador2024}. In these studies, the rephasing signal (defined in~\cite{Mahmood2021}) along $\omega_\tau = -\omega_t$ in the 2D spectrum exhibits different line widths along the diagonal and off-diagonal cuts.}
\textcolor{black}{However, in the case of \mgb, the nonlinear 2D spectrum does not show the rephasing signal precisely along $\omega_\tau = -\omega_t$ line in the 2D spectrum, as displayed in Fig.~\ref{fig_broad}. This is very similar to that in NbN~\cite{Katsumi_NbN_2023}. Therefore, we conclude that the current 2D spectrum does not allow us to easily distinguish the various broadening mechanisms.}

\begin{figure}[t]
	\centering
	\includegraphics[width=\columnwidth]{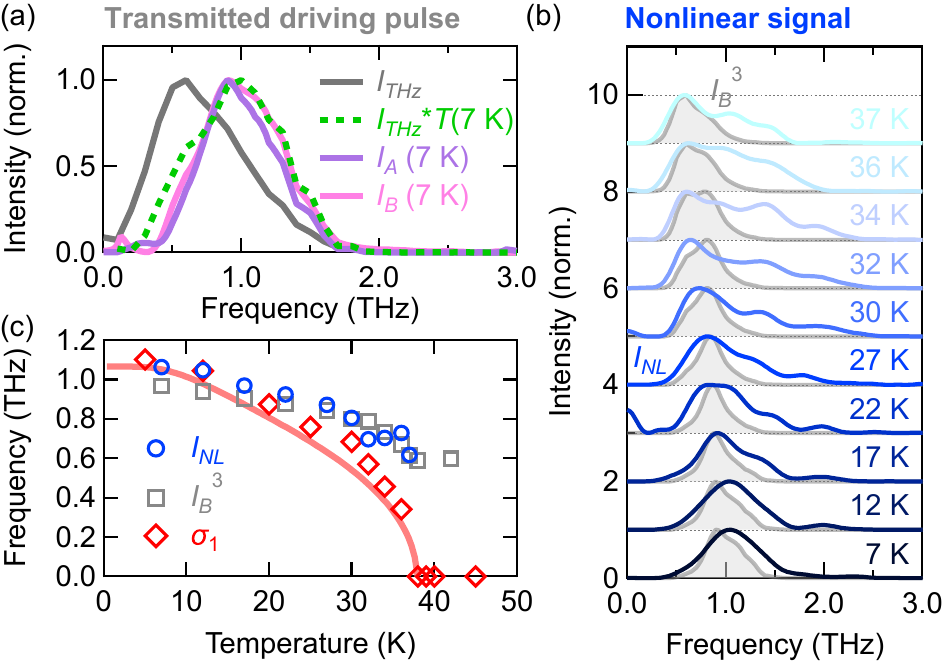}
	\caption{\textcolor{black}{(a) Power spectrum of the THz field measured at the sample position (gray) and that multiplied by the power transmission coefficient of \mgb at \textcolor{black}{7}~K (green). The power spectra of the $A$ and $B$ pulse's THz field transmitted through the sample at 7~K are shown by the purple and magenta curves, respectively. (b) Power spectrum of the nonlinear signal ($I_{NL}$) at $\tau = 0$~ps at selected temperatures. The cube of the transmitted $B$-pulse's intensity ($I_B^3$) are also shown by gray curves. (c) Temperature dependence of the peak energy of $I_{NL}$ (blue) and $I_B^3$ (gray). The red diamond is twice the SC gap $2\Delta_{\pi}/2\pi$ obtained from the optical conductivity in Fig. S1. The red curve is the solution of the BCS gap equations.}}
	\label{fig_broad_5K}
\end{figure}

\textcolor{black}{\section{Transmitted THz driving fields for the broad-band THz 2DCS}}
\textcolor{black}{
For the broad-band THz driving case, we identified a peak in the nonlinear signal's power spectra $I_{NL}$ at \pdelta~=~1~THz at 7~K. Since neither the THz driving field nor the transmission coefficient of \mgb show a peak at 1~THz, the peak in $I_{NL}$ was attributed to the resonance of the $\pi$-band amplitude mode. To further confirm this interpretation, we consider the transmission coefficient of \mgb. When the THz driving field's spectra are multiplied by the transmission coefficient, peaks appear around 1 THz, as shown in Fig.~\ref{fig_broad_5K}(a). This coincides with the transmitted $A$ and $B$-pulse's power spectra shown by the purple and magenta curves, respectively. Since the nonlinear signal arises from a third order process, we compare the temperature dependence of the nonlinear signal and cube of the $B$-pulse power spectra in Fig. S7(b). The peak positions of $I_{NL}$ and $I_B^3$ are slightly different below 20~K indicating that the nonlinear susceptibility has some frequency structure. Their peak energy is evaluated by Lorentzian fitting and summarized in Fig.~S7(c). Below 20~K, the peak in $I_{NL}$ matches with \pdelta (red) obtained from the optical conductivity and above 20~K approaches the peak in $I_B^3$. \textcolor{black}{The deviation of the peak in $I_{NL}$ from \pdelta above 20~K suggests that the amplitude mode is overdamped, which is likely due to the interband coupling.} Furthermore, the width of $I_{NL}$ is wider than that of $I_B^3$. We believe the differences in peak energy and width indicate that the peak in $I_{NL}$ below 20~K arises from the resonance of the $\pi$-band amplitude mode.
}

\begin{figure}[t]
	\centering
	\includegraphics[width=4.5cm]{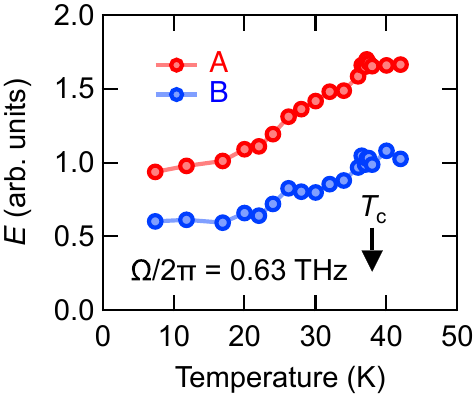}
	\caption{Transmitted $E$-field amplitude at $\Omega/2\pi=0.63$~THz of the $A$ and $B$ pulses as a function of temperature. Here, the field strength of the $A$ and $B$ pulses are set to 12~kV/cm.}
	\label{fig_a_b_temp}
\end{figure}

\ 
\section{Normalization of the nonlinear signals by measured THz fields}

\begin{figure*}[t]
	\centering
	\includegraphics[width=14cm]{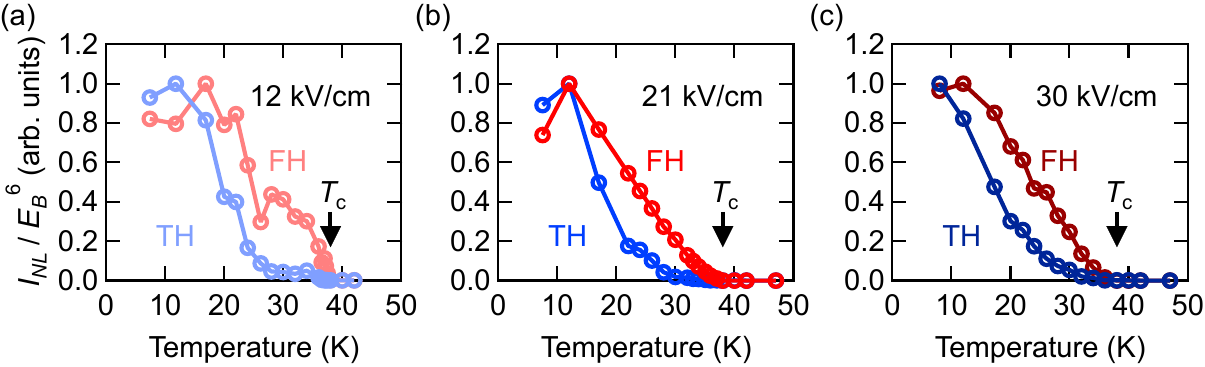}
	\caption{\textcolor{black}{Temperature dependence of the nonlinear signal's intensity for the FH and TH normalized by the sixth power of the $B$-pulse's amplitude $E_B^6$ for the field strength of (a)12, (b) \textcolor{black}{21}, and (c) 30 \kv. All the signals are further normalized by their maximum values for comparison.  The increasing intensity of red and blue colors correspond to the driving THz intensity.}}
	\label{fig_norm}
\end{figure*}

Figure~\ref{fig_a_b_temp} shows the transmitted $E$-field strength of the $A$ and $B$ pulses at 0.63~THz as a function of temperature. Due to the temperature-dependent change in the refractive index, the transmitted and internal $E$-field decreases when the temperature is lowered. This screening effect must be considered to evaluate the temperature dependence of the third-order nonlinear susceptibility, as documented in Ref.~\cite{Katsumi_NbN_2023}. We normalized the first-harmonic (FH) and third-harmonic (TH) signals by the sixth powers of the $E$-field of the $A$-pulse in Fig.~4 in the main text. While the raw data of the FH signal in Fig.~3(d) for 12~\kv displays a enhancement around 35~K, the normalized FH intensity shows a monotonic increase toward lower temperature. We note that the normalized results do not depend largely on which pulse is used for normalization. Figure~\ref{fig_norm}\textcolor{black}{(a)} presents the intensity of the FH and TH signals as a function of temperature normalized by the six powers of the $E$-field of the $B$-pulse, consistent with the results in Fig.~4. \textcolor{black}{We note that the normalized nonlinear signals with the $B$-pulse are noisier than those with the $A$-pulse due to the imperfect focusing of the $B$-pulse to the detection position, making the detected $B$-pulse amplitude about two times smaller than the $A$-pulse one. This is why we employed the $A$ pulse for the normalization in the main text. We confirmed the monotonic increase of the normalized FH and TH signals toward lower temperatures for the field strength of \textcolor{black}{21} and 30~\kv, as shown in Figs. \ref{fig_norm}(b) and (c).} These results indicate that the resonant-like enhancement observed in the raw data of the FH intensity is simply due to the temperature-dependent refractive index.

\bibliographystyle{apsrev4-1}
\bibliography{MgB2.bib}